# From Dust to Planets – A Chemical Perspective

Klaus Mezger[1*], Jonas Pape[1,2], Aryavart Anand[1,3], Pascal M. Kruttasch[1], Hauke Vollstaedt[1,4], Jan Hoffmann[1]

[1]Institut für Geologie, Universität Bern, Baltzerstrasse 1+3, 3012 Bern, Switzerland

[2]Institut für Planetologie, Universität Münster, Wilhelm-Klemm-Straße 10, 48149 Münster, Germany

[3]Max Planck Institute for Solar System Research, Justus-von-Liebig-Weg 3, 37077 Göttingen, Germany

[4]Thermo Fisher Scientific, Hanna-Kunath-Straße 11, 28199 Bremen, Germany

[*]Email: klaus.mezger@unibe.ch

# Abstract

Chemical and chronological information preserved in meteorites permits the reconstruction of events and processes in the solar nebula from the formation of the first solids to the accretion of planetary bodies and their subsequent differentiation. The path from a gas-dust cloud to differentiated planets includes intervals of steady evolution interrupted by singular events that dramatically altered this steady path, leading to planetary bodies with distinct chemical compositions and different degrees of internal differentiation. The dominant continuous process in the early Solar System was the cooling of the gas-dust cloud, which caused a steady condensation of elements into solid compounds and a continuous increase in the dust/gas ratio. Planetesimal formation started within less than 1 Ma of Solar System formation and continued for ca. 3 Ma apparently in random regions within the disk. The first planetesimals most likely formed due to streaming instabilities and created gaps in the gas-dust disk that prevented significant element exchange. Later planetesimals formed by accretion of chondrules that had developed in the dust rings by bow shocks. The Earth formed by early accretion of volatile-poor material and a later collision with a Mars-sized volatile richer body after proto-Earth had formed a metal core. This chance event provided the chemical conditions that transformed the Earth into a habitable planet.

# Introduction

A major current interest of planetary science is to discover the diversity of planets and planetary systems and then target potentially habitable planets for detailed investigation. A guided search can be assisted by the knowledge about the physical and chemical processes that led from a molecular cloud to a planetary system with potentially habitable planets. One goal in this endeavor is the construction of a unified or general physical model that describes the processes from the collapse of a molecular cloud to the formation of a planetary system and also satisfies chemical constraints. Our Solar System may represent one of multiple possible pathways from the formation of the first solids to the completion of a multi-planetary system that may contain planets with habitable worlds on which life has emerged.

On the path toward the development of a general theory of planet formation that describes the origin and evolution of diverse planetary systems (e.g., Mordasini and Burn, 2024), it could be of great help to understand in some detail the origin of our Solar System and its evolution in space and time. For exoplanets only a very limited data set on their physical parameters and chemical compositions can be obtained with current technologies. With space telescopes it is possible to determine physical parameters like mass, radius, and orbital location. Information on the chemical composition of exoplanets is generally only available for their atmospheres (e.g., Kitzmann et al., 2025). Abundances of heavy elements can only be determined spectroscopically from very hot exoplanets, but these cannot be habitable. Detailed chemical, isotopic and mineralogical information is currently not obtainable for potentially habitable exoplanets.

Studying the Solar System in detail has the advantage that a wide diversity of information on different planetary bodies (asteroids, planets, moons) is accessible via space missions that use remote sensing to analyze the morphology, mineralogy and chemical composition of the surface. Some physical and chemical data are available from in-situ measurements on planetary bodies as part of space missions (e.g., Moon, Venus, Mars, asteroid 67P/Churyumov–Gerasimenko). However, high-precision and high accuracy chemical compositions can only be obtained from rocky materials that originate from different regions of the Solar System and can be analyzed in a laboratory with state-of-the-art analytical facilities. Materials that can be studied directly in the laboratory with multiple analytical methods include around 73,000 meteorites to date (Meteoritical Bulletin Database), representing a multitude of planetary bodies of different sizes and compositions. These meteorite samples are augmented by micrometeorites, which are, however, strongly modified during their passage through the Earth`s atmosphere. In addition, samples returned by space missions from the Moon, primitive planetesimals (e.g., Ryugu, Bennu) (Grady et al., 2025) and tiny amounts of cometary dust collected by the Stardust mission (Brownlee, 2014) are available for direct investigation.

The chemical, isotopic and mineralogical compositions of meteorites and rock samples returned form asteroids reflect the physical conditions and chemical processes of their formation during the early stages of the evolution of the protoplanetary disk, and the diversity of the different planet-building materials. The diverse meteorites represent different arrested stages from the first formation of solids in the solar nebula to differentiated planets. These samples allow direct analysis of their chemical, isotopic and physical properties and thus provide measurable information on their origin and chemical evolution, including the absolute time of these events and processes. Meteorites and their different components are invaluable for understanding the origin and evolution of material in the Solar System and provide key constraints on the origin of the elements and the evolution of a planetary system.

This Chapter reviews key research results obtained on meteorites and their components, made possible with support from the NCCR PlanetS. During the course of these research projects interactions with PlanetS members and their interdisciplinary expertise were instrumental in conceiving and developing the research projects and placing the results in a larger context of planet formation.

# 1 Composition of the solar nebula

Nucleosynthetic isotope heterogeneities of a variety of elements in rocky samples of the Solar System are inherited from the solar nebula and indicate that the gas-dust cloud from which planetesimals ultimately formed was heterogeneous during the formation of solids (e.g., Burkhardt et al., 2019; Clayton, 1982; Cook and Schönbächler, 2017; Cook et al., 2021; Dauphas et al., 2002; Ek et al., 2020; Nanne et al., 2019; Rüfenacht et al., 2023; Schneider et al., 2023; Yap and Tissot, 2023). These mass-independent stable isotope variations among meteorites from different planetesimals, Mars and terrestrial samples indicate that the chemical elements of the Solar System originate from multiple nucleosynthetic sources with different paths of nucleosynthesis. Thus, the present-day Solar System materials are the product of different nucleosynthesis processes and pathways in different stars. These nucleosynthetic stable isotope anomalies record information on mixing and fractionation processes in the early solar nebula (e.g., Warren, 2011a; Burkhardt et al., 2019; Spitzer et al., 2022; Palme and Mezger, 2024).

The observed isotope variations imply that the solar nebula was fed from multiple nucleosynthetic sources and was heterogeneous prior to accretion of solids into planetesimals, external material was added during the formation of solids or variations in the cloud were generated by thermal processing of solids and/or size sorting (e.g., Hutchison et al., 2025). Thus, the isotope variability of meteorites is a proxy for space and/or time in the evolution of the earliest Solar System (Mezger et al., 2020). In combination with bulk chemical compositions of different meteorite classes that represent different planetesimals, it is possible to reconstruct parts of the physical and chemical processes in the early Solar System leading up to planetesimal accretion. Particularly the primitive chondritic meteorites provide information on mixing and fractionation processes for the region in the solar nebula where they formed.

A compilation of nucleosynthetic isotope anomalies for multi-isotope elements from available Solar System materials shows a highly systematic relationship among the different isotope anomalies for many elements (e.g., Warren 2011a; Burkardt et al., 2019; Ek et al., 2020; Rüfenacht et al., 2023). In a two-isotope diagram (e.g., $^{54}$Cr vs. $^{50}$Ti) the samples from the non-carbonaceous chondrite (NC) reservoir (which includes ordinary chondrites (OC), enstatite chondrites (EC), HED meteorites, ureilites, NC irons meteorites, Earth, Moon, and Mars) define one end of the array. Ca-Al-rich inclusions (CAI) define the other end and the samples from the carbonaceous chondrite (CC) reservoir plot between the endmembers (Figure 1) (see compilations in Burkhardt et al., 2019; Rüfenacht et al., 2023; Dauphas et al., 2024). These systematic correlations point to two distinct nucleosynthesis environments that were the dominant sources of the elements that made up the solar nebula. It also demonstrates that the materials from the two main nucleosynthetic sources were not homogenized in the solar nebula prior to the formation of planetesimals. In addition to these two major nucleosynthesis sources minor contributions are documented by ureilites (e.g., Kruttasch et al., 2025) and refractory grains in primitive meteorites (e.g., Hoppe 2008).

Combining the relative abundances of neutron-rich isotopes (e.g., $^{48}$Ca, $^{50}$Ti, $^{54}$Cr) from primitive and differentiated meteorites with Earth, Moon and meteorite samples from Mars yields a well-defined systematic array in two-isotope space (e.g., Burkhardt et al., 2019; Rüfenacht et al., 2023) (Figure 1). Materials from the NC reservoir have the lowest abundance of neutron-rich isotopes. Meteorites from the CC-reservoir have higher abundances and the highly refractory Ca-Al-rich inclusions (CAI) in different meteorites have the highest abundance of neutron-rich isotopes. For refractory elements the two-isotope plot defines a nearly perfect linear array (e.g., Burkhardt et al., 2019; Palme and Mezger, 2024). Deviations from this array are observed for CCs when the moderately volatile elements, like Cr, are included (e.g.,Trinquier et al., 2009).

In $\varepsilon^{54}$Cr versus $\varepsilon^{50}$Ti space, OC, Earth, EC, Mars, CI and CAIs are highly correlated ($R^2>$ 0.99, Palme and Mezger 2024) (Figure 1), but CCs are significantly above the regression line; while ureilites and the meteorites from the Vesta-like subgroup (Rüfenacht et al., 2023; Kruttasch et al., 2025) are slightly below the regression line. This linear array, the *Chondrite Reference Line* (CRL), can be interpreted as a two-component mixing line. In order to define a straight mixing line in $\varepsilon^{50}$Ti-$\varepsilon^{54}$Cr (Figure 1), the Cr/Ti elemental ratio needs to be the same in both endmembers. The well-defined linear relationship indicates that all meteorite reservoirs along the CRL began with a CI-chondritic (solar) Cr/Ti ratio, regardless of their present Cr/Ti ratios. Materials with defined element abundances are minerals as they possess a defined stoichiometry. This implies that this mixture of material from two different nucleosynthetic sources is the result of mixing solid crystalline dust. Thus, this array suggests that Cr and Ti were present in the same types of refractory minerals, but these minerals originated from distinct stellar sources and had distinct isotope compositions. The well-defined correlation line also implies that the solar nebula derived from at least two distinct supernova sources with different nucleosynthesis pathways. Mixing of the two major components must have been an early process, most likely at the time when the solar nebula was still a gas-dust cloud.

The preservation of isotope variability among different Solar System material indicates that the cloud was not homogenized or became unmixed (e.g., Hutchinson et al., 2022, 2025). However, creating this array by unmixing is difficult, if the elements that have different isotope composition reside in the same refractory mineral. The endmembers that define the *CRL* include one component with low abundance of neutron-rich isotopes and chondritic abundance of refractory and moderately volatile elements and a second component that also had a chondritic composition, but the isotope abundances now recorded in CAIs ("IC component": Burkhardt et al., 2019; "CAI-like dust": Yap and Tissot, 2023). As shown in Figure 1, the carbonaceous chondrites deviate from the CRL. This is due to their later contamination with CAI material indicated by the mixing lines in the figure. This CAI material was strongly depleted in volatile elements, but this volatile loss did not affect the isotope composition. Subtracting this secondary CAI component from carbonaceous chondrites would put the data points on the CRL in $\varepsilon^{50}$Ti versus $\varepsilon^{54}$Cr space (see Trinquier et al., 2009). Since CAIs are extremely depleted in volatile elements, they must have been modified by a high temperature process. This depletion occurred probably close to the Sun. Subsequently this material was transported outward into the region of the CC reservoir. This process requires the transfer of solids from the inner towards the outer Solar System prior to accretion of the carbonaceous chondrite parent bodies.

The combination of isotope anomalies with the composition of different solid materials from the Solar System indicates that mixing of different materials occurred at distinct stages during the early evolution of the solar nebula. A first major mixing event occurred when dust from two distinct nucleosynthetic sources was mixed. These dusts had similar element abundances but distinct isotope composition. A second significant mixing event occurred when material from the inner Solar System that was highly depleted by thermal processing closer to the Sun, i.e., CAI material up to pebble size, was transported into the region where carbonaceous chondrites formed.

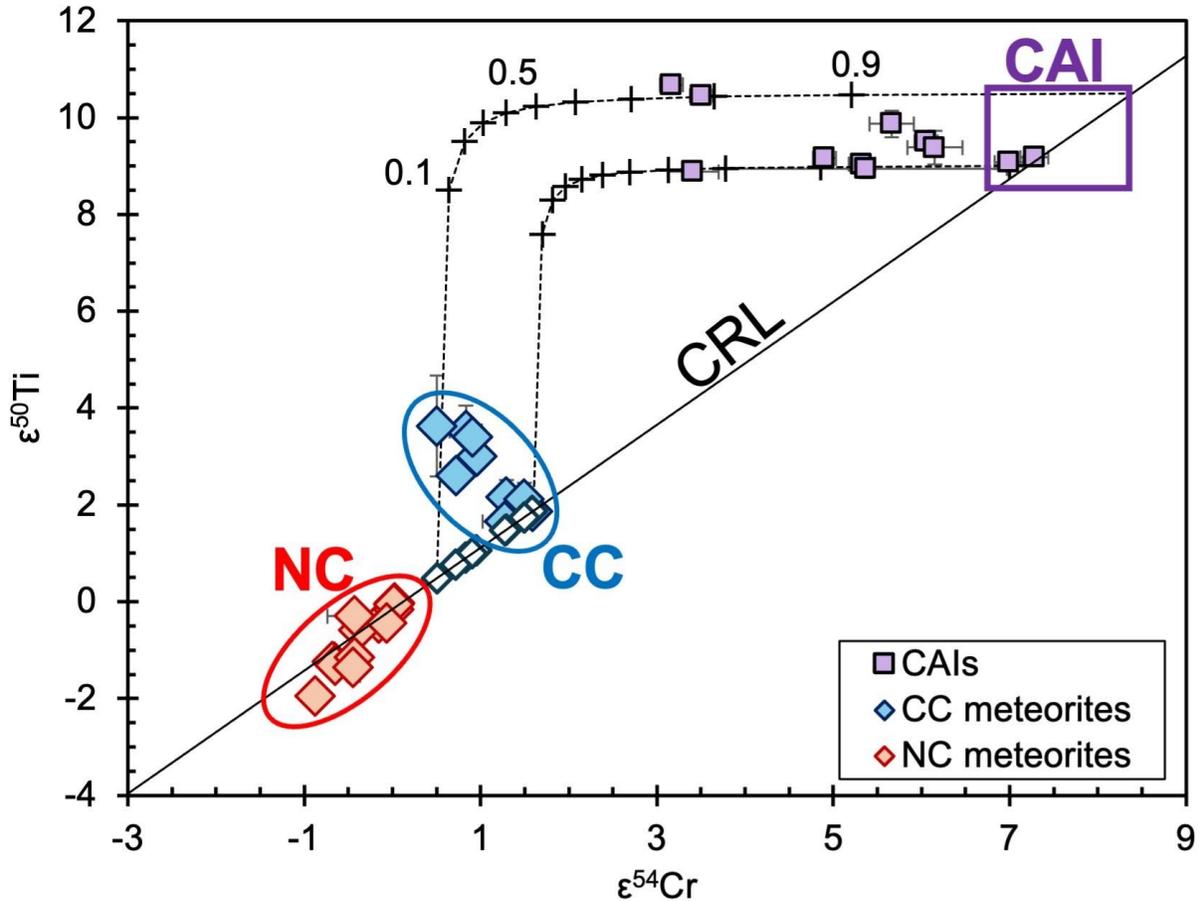

Figure 1: The isotope variability in meteorites from different sources record two major mixing events. The *Chondrite Reference Line* (CRL) defines the mixture of material from different nucleosynthetic sources but with similar element abundances (i.e., first mixing event). A second mixing event is recorded by meteorites from the CC-reservoir that obtained volatile depleted material similar to CAIs. Removal of this CAI component places the carbonaceous chondrites on the CRL (open diamonds). Numbers on a mixing curve give the fraction of CAI component added. Data compiled from literature (e.g., Palme and Mezger, 2024) (NC = non-carbonaceous chondrite reservoir, CC = carbonaceous chondrite reservoir, CAI = Ca-Al-rich inclusions).

## 2 Condensation of the elements from the solar nebula

When the element abundances of solids from the Solar System are plotted versus the condensation temperature of the element (Lodders 2003; Wood et al., 2019) all these materials share a common feature, i.e., relative chondritic abundances of refractory elements and some degree of depletion in volatile elements. The extent of depletion correlates with the condensation temperatures of the elements. All meteorite groups show this depletion but to very different degrees. The least depleted samples are CI chondrites, which are missing only highly volatile elements; all the other elements have abundances very similar to the Sun and thus the bulk Solar System. The strongest depleted materials are CAIs, which consist of dominantly highly refractory elements, particularly the abundant elements Ca and Al.

The element depletion is highly systematic in the group of carbonaceous chondrites (e.g., Braukmüller et al., 2018, 2019; Vollstaedt et al., 2020). In these samples the degree of depletion is strongly dependent on the condensation temperature of the elements (Figure 2). This relationship indicates that the refractory elements condensed completely into minerals up to the temperature range when Si-Fe-Mg condensed from the cooling nebula. Elements with condensation temperature lower than Si-Fe-Mg (moderately volatile elements) show a systematic depletion trend that is characteristic for each group of carbonaceous chondrites. This depletion is commonly attributed to partial condensation from a cooling solar nebula. However, this trend is not what is expected from partial condensation. As almost all elements condense to more than 95 % over an interval of less than 50 °C (Lodders et al., 2025), a step pattern would be expected with refractory elements condensed completely and volatile element did not condense. Such patterns are recorded in HED meteorites that derive from the planetesimal 4Vesta (see Figure 1 in Sossi et al., 2022) or inferred for proto-Earth (Mezger et al., 2021).

An important consequence of partial condensation from a hot nebula is the mass-dependent fractionation of isotopes that correlates with the degree of depletion. It is then expected that the variations in the stable isotope composition of an element should correlate with the degree of condensation or the degree of element depletion relative to the bulk Solar System, with the material with the strongest depletion showing enrichment in heavy over light isotopes. The opposite trend is observed for some elements (e.g., Morton et al., 2024; Wölfers et al., 2025). Thus, these stable isotope variation trends are most likely due to mixing of a volatile-rich and isotopically heavy component with one that is volatile-poor and isotopically light (e.g., Morton et al., 2024; Wölfers et al., 2025). Therefore, the correlation of isotope composition with element depletion cannot be attributed to partial condensation or partial evaporation during formation of solids from the nebula.

The depletion of the moderately volatile elements without an enrichment of heavy isotopes is thus most likely due to a physical rather than a chemical process.

The correlation of element abundances with the condensation temperature, but not with concomitant stable isotope fractionation, requires an alternative model to partial condensation in a cooling solar nebula. A process is needed that removed gas from the nebula, but not the condensed solid, and this way changed the surface density of the solar nebula as a function of time and temperature. The gradual decrease of element abundances, relative to CI composition, with condensation temperature is the consequence of removal of volatile material from the surface of the cooling nebula and the settlement of the solids towards the mid-plane of the solar nebula (Vollstaedt et al., 2020; Sengupta et al., 2022). The mechanical removal of residual gas from the nebula can account for the depletion of the remaining volatile elements without mass-dependent isotope fractionation.

During cooling of the nebular cloud, the amount of solids depends on the condensation temperature of the elements and their abundance in the cloud. The mass of solid material with condensation temperatures above 1400 K amounts to less than 5% of the condensable elements (Figure 2). Between 1400 and 1300 K Si-Fe-Mg condense, which are the most abundant metallic elements in the Solar System. Condensation of these elements results in a strong increase in solids. The consequence is a dramatic change in the gas-dust ratio in the solar nebula leaving behind a less opaque gas nebula whose transparency is enhanced by the migration of the solids to the mid-plane, where they may coagulate into larger masses, possibly even planetesimals, that are devoid of volatile elements. Interaction of the remaining gas with the radiation from the Sun leads to mechanical removal of gas from the surface of the cloud. In their quantitative model for volatile element depletion Sengupta et al. (2022) modelled mass loss driven by disk winds to drive nebular cooling during condensation. Vollstaedt et al. (2020) suggested mass-loss by photo-evaporation. Condensation followed by coagulation of solids from a cooling nebula of diminishing mass resulted in the observed pattern of decreasing volatile abundances with decreasing condensation temperature. Thus, the depletion of an element relative to the refractory elements is not due to incomplete condensation, but rather due to the separation of more volatile material from condensed material as the nebula cools over time. The slopes of the depletion patterns for the different carbonaceous chondrites are consistent with independent constraints on the solar nebula cooling times and the decrease of the surface density (Cassen, 1996; Humayun and Cassen, 2020). The different slopes correspond to differences in the degree of moderately volatile element depletion for the source regions of the different chondrite parent bodies. The degree of depletion of the carbonaceous chondrites shows a correlation with nucleosynthetic anomalies (Figures 1, 2). The most volatile depleted CV meteorites have the lowest and the undepleted CI meteorites have the highest abundance of neutron-rich isotopes. Thus, the composition of carbonaceous chondrites may correlate with the heliocentric distance of their respective feeding zones.

A characteristic feature of all element patterns for carbonaceous chondrites is their close to chondritic relative abundance of volatile elements ("hockey stick pattern"; Figure 2) that is supported by high quality measurements of these elements (Braukmüller et al., 2019). This change in the element trend implies a cessation of the mechanical removal of volatile material and a sudden condensation of the remaining elements in the cloud (Neuland et al., 2021). This sudden change in condensation behaviour may be related to the formation of chondrules which significantly decreased nebula opacity. As a consequence, the nebula cooling rates increased, leading to a nearly quantitative condensation of highly volatile elements in chondritic relative abundance (Vollstaedt et al., 2020).

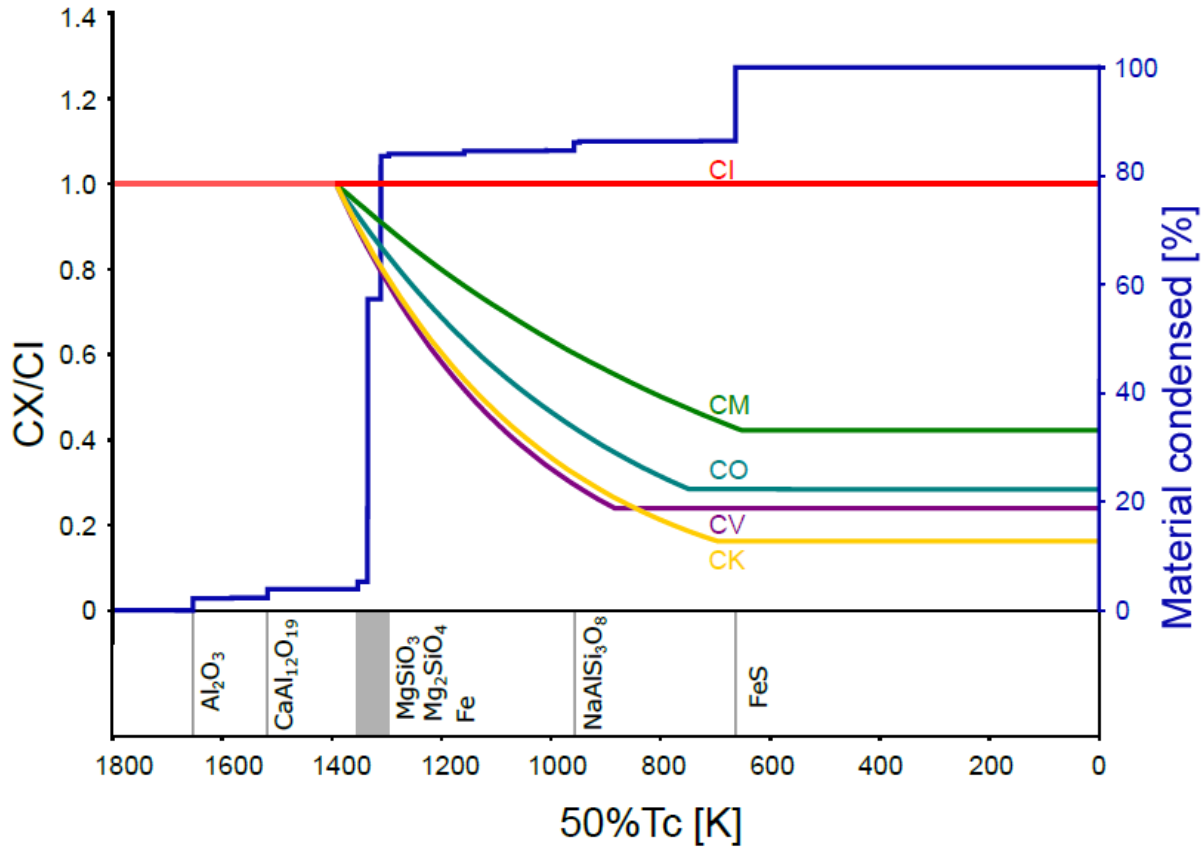

Figure 2: Regressions of CI-normalized elements concentrations in CM, CO, CK, and CV chondrites. Carbonaceous chondrites show different extents of moderately volatile element depletion, but relative chondritic abundances of refractory and volatile elements. The amount of solids increases dramatically when the highly abundant elements Si, Fe and Mg condense as silicates of metal (modified after Vollstaedt et al., 2020)

## 3 Chondrules and Chondrites

Among all known meteorites the CI chondrites have a chemical composition that is closest to that of the Sun and thus the bulk Solar System. These meteorites are, however, extremely rare among known samples. The meteorite collection has been augmented by CI material collected on the asteroids Ryugu and Bennu (e.g., Barnes et al., 2025, Yokoyama et al., 2023). The samples with the next closest composition to CI chondrites are the chondrule-bearing chondrites. Thus, chondrites represent the most primitive material from the Solar System, and they have been considered as the primary building blocks of planetesimals and planets, including the Earth.

Chondrites are the most abundant type of meteorites in the collections, making up ca. 90% of all meteorite samples (Meteoritical Bulletin Database). There are three major classes of chondrites: ordinary, enstatite and carbonaceous chondrites. Each class is divided into several groups and subgroups (Krot et al. 2014; Scott and Krot 2014). There are additional groups that do not belong to anyone of these classes (e.g., Rumuruti-like (R) or Kakangari-like (K) chondrites), as well as individual anomalous chondrites that are rare (< 5 individuals) and are not placed in a named group. The common characteristic of these chondrites is the presence of (sub-)millimeter-sized spherical mineral parageneses, i.e., chondrules, in different abundances ranging from a few vol% to ca. 90% (e.g., Palme et al., 2015). The chondrules are surrounded by a fine-grained mineral matrix. Chondrules contain unambiguous evidence of having been melts (e.g., Jones, 2024) and thus are the product of a high-temperature event. This melting process is evident from their mineral textures and the presence of glass in highly pristine chondrules (metamorphic grade 3). In addition to chondrules, particularly the carbonaceous chondrites also contain CAIs. These highly refractory inclusions are very rare or even absent in ordinary and enstatite chondrites which belong to the NC-group (see Figure 1) (e.g., Scott and Krot, 2014). All chondrites (except for CI) contain metal (Fe-Ni alloy) and commonly the Fe-sulfide troilite.

Since chondrites are the dominant materials available for direct study in the laboratory, they are likely of great significance for reconstructing early Solar System processes from the formation of the first dust to the accretion of planetesimal. However, the chondrule forming process has remained quite enigmatic. For the formation of chondrules different processes have been proposed: 1) planetesimal collisions (e.g., Lichtenberg et al., 2018), 2) lightning in the early Solar System (Horanyi et al., 1995), or 3) bow shocks due to the migration of planetesimals through the solar disk (e.g., Morris and Boley, 2018). Each of these proposed mechanisms has its shortcomings. A collision of two differentiated planets may not be able to homogenize all the material enough to achieve chondritic bulk compositions again. Lightning may have been only a local phenomenon. Bodénan et al. (2020) modelled bow shock and found that chondrule formation is apparently only possible in a very narrow range of physical conditions. They conclude that it "seems unlikely that shocks from Jupiter can form chondrules in most cases". Precise ages for the formation of chondrules, particularly the formation of melt in individual chondrules can provide key information on the chondrule forming process.

Dating early Solar System processes and events with high precision is possible through the application of short-lived isotope systems. In the case of chondrules the $^{26}$Al-$^{26}$Mg decay system is well suited, if the different components in chondrules have very different Al/Mg ratios. The dominant silicates in chondrules are olivine and orthopyroxene, both have Al/Mg ratios near zero. The high Al/Mg phase in chondrules is glass (or mesostasis if recrystallized), or in rare examples plagioclase. Thus, individual chondrules with large enough parts of glass or mesostasis are needed to obtain high precision ages with in-situ dating methods. The most commonly used method is secondary ion mass spectrometry (SIMS). Figure 3 shows a compilation of published $^{26}$Al-$^{26}$Mg ages of pristine chondrules from ordinary and carbonaceous chondrites (Hutcheon and Hutchison, 1989; Kita et al., 2000; Mostefaoui et al., 2002; Kita et al., 2005; Rudraswami and Goswami, 2007; Rudraswami et al., 2008; Villeneuve et al., 2009; Mishra and Goswami, 2014; Pape et al., 2019; Siron et al., 2021, 2022; Yurimoto and Wasson, 2002; Kunihiro et al., 2004; Sugiura and Krot, 2007; Kurahashi et al., 2008; Hutcheon et al., 2009; Ushikubo et al., 2013; Nagashima et al., 2017; Hertwig et al., 2019; Fukada et al., 2022; Piralla et al., 2023). Data for enstatite chondrites are rare due to the general absence of glass or pristine mesostasis. For the compilation of chondrule ages only data from pristine meteorites are considered (i.e., metamorphic grade ≤3.15), and where the age uncertainty is less than $\pm$1.5 Ma (2$\sigma$) for individual isochrons.

The compilation shows that chondrules in carbonaceous and ordinary chondrites formed at similar times and over similar time scales but show well-defined age peaks (Figure 3). The major chondrule forming event occurred at ca. 2.0 Ma in the OC chondrite reservoir and ca. 0.5 Ma later in the CC reservoir. Chondrules in CR meteorites formed later between 3-4 Ma after CAIs (Schrader et al., 2017; Tenner et al., 2019; Nagashima et al., 2014). Within a single sample the range in ages of individual chondrules is similar within analytical uncertainty pointing to a singular chondrule-forming event. However, some rare chondrules show chemical and petrographic evidence for partial remelting due to a second thermal pulse (Pape et al., 2021). This second thermal pulse must have affected the chondrule prior to their incorporation into the chondrite parent body. These complex chondrules indicate that chondrule formation occurred as multiple punctuated thermal events that achieved temperatures significantly above the solidus of the mineral assemblage in some chondrules and occurred in the same region of the solar nebula. The distinct chondrule forming episodes (OC, CC, CR chondrules) suggest that the chondrule forming process was not contemporaneous throughout the solar nebula, but occurred multiple times and individual events were limited to specific regions of the gas-dust disk (Fukuda et al., 2022).

Bulk samples from different meteorites groups have distinct variation in nucleosynthetic anomalies (Figure 1). Similar, but much smaller variations are recorded in different individual chondrules from the same meteorite sample (Rüfenacht, 2022; Marrocchi et al., 2022; Anand and Mezger, 2025). This implies that the chondrules from a given meteorite group are sourced from a distinct and limited region of the solar nebula and there was only minimal exchange of chondrules between these regions. However, some carbonaceous chondrites, including CV and CK meteorites, contain large chondrules from the NC-reservoir (Williams et al., 2020; van Kooten et al., 2021). Thus, some outward transport of material across gaps was possible and these gaps were not completely impermeable to chondrule migration.

Accretion of a first generation of planetary bodies led to a separation of dust regions from each other which became the dominant source for the chondrules of individual planetesimals. The chondrules formed in these dust rings and accreted to individual chondrite parent bodies. Planetesimals or planetary embryos in the disk separated regions of dispersed dust and gas and could have been an efficient source for bow shocks that caused chondrule formation by melting of pre-existing dust agglomerates. This scenario may be documented by the observed ring systems around nascent planetary systems (e.g., Andrews et al., 2018). This model is consistent with the chemically primitive composition of chondrites and the formation of chondrules in different regions following the accretion of the first planetary bodies. Thus, chondrules and chondrites are not building material of the first planetesimals, but rather the result of early planetesimal formation. In this case it is likely that the first planetesimals formed due to streaming instability in the solar nebula (e.g., Drążkowska and Alibert, 2017; Nesvorný et al., 2019; Lim et al., 2024; Helled et al., 2025). Thus, some planetary bodies formed early, and the remaining dust rings preserved a chemical primitive bulk composition up to ca. 3 Ma after the formation of the first differentiated bodies. This is consistent with $^{53}$Cr model ages of carbonaceous chondrites, which date their last Mn-Cr fractionation event from the bulk Solar System (Kruttasch and Mezger, 2025). Due to the already low abundance of $^{26}$Al at the time of chondrite parent bodies, these planetesimals did not differentiate into a metallic core and silicate mantle. They preserved their primary textures and chemical composition, giving them the appearance of primitive Solar System material, despite their late accretion into chondrite parent bodies.

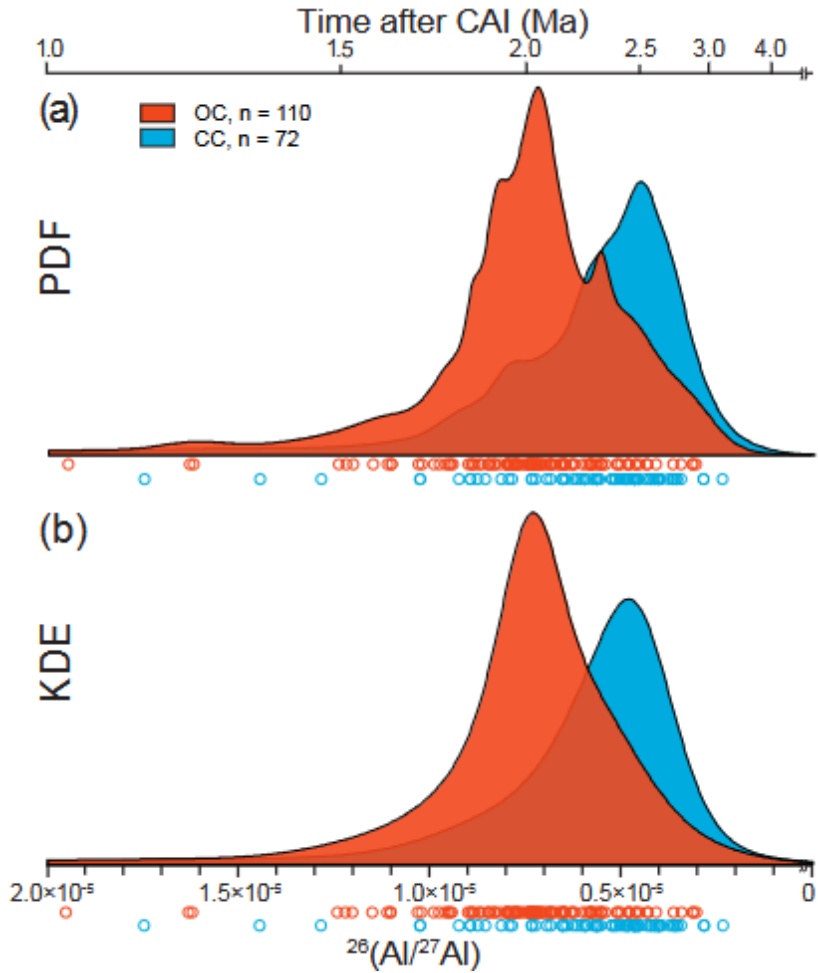

Figure 3: Compilation of published chondrule ages determined with the $^{26}$Al-$^{26}$Mg chronometer. The peaks of chondrule formation differ between OC and CC by ~0.5 Ma. Chondrules in ordinary chondrites are only from samples with a metamorphic grade ≤3.15. Only samples with a 2σ uncertainty <±1.5 Ma are considered. PDF: Probability density function; KDE: Kernel density function. (See text for data sources.)

## 4 Planetesimal accretion and differentiation

A still enigmatic process is the formation of planetesimals from dust. As the solar nebula cooled the elements condensed into minerals forming solids in the form of dust-sized mineral grains (Section 2). The extent of dust formation depends on the thermal structure of the nebula and its cooling history. During cooling a maximum amount of solids is formed around the temperature interval where Fe-Si-Mg condense as metal and silicates (Figure 2). These three elements, together with O, are the most abundant elements in rocky planets. The most abundant elements in the solar nebula, He and H remained as gas in the outer regions. Runaway gas accretion onto a core lead to the formation of gas planets (e.g., Alibert et al., 2018). The solid particles collided and stuck together and eventually grew to planetesimals. For the path from dust to km-sized rocky planetesimal two major models have been proposed: a) planetesimal formation due to streaming instability in the region of the mid-plane of the solar nebula where the highest dust concentration occurred (e.g., Lim et al., 2024, Schäfer et al., 2017) , or, b) planetesimal formation by pebble accretion, where mm- to cm-sized solids combine rapidly into planetesimals or planets (e.g., Johansen and Lambrechts, 2017).

It is not possible to obtain radiometric ages directly for the accretion process. Ages can only be obtained for chemical fractionation processes that occurred as a consequence of planetesimal accretion. These fractionation processes are controlled by heating of the accreting body. In the early Solar System, the two major heat sources were collisions of bodies and the decay of the short-lived isotope $^{26}$Al. Since $^{26}$Al has a short half-life of 0.717 Ma (National Nuclear Data Center, NuDat v3, 2023), radioactive decay is a significant heat source only during the first ~3 Ma. Afterwards, collisions are the sole significant heat source for planetary bodies capable of inducing melting and planetwide differentiation. Heating of an accreting body leads to metamorphism and thus new minerals grow that can be used to date the time of their formation. Further heating of planetesimals can lead to the formation of a metal, silicate or a sulfide melt. Segregation of a metal or sulfide melt leaves behind a silicate mantle. Due to the density differences the heavy metal or sulfide melt segregates toward the center of the body forming a core and a silicate mantle. Ages for this process can be obtained either from the metallic core or the silicate mantle material derived from differentiated planetesimals and available in the form of meteorites (e.g., Kleine et al., 2005; Anand et al., 2021a, b).

The formation of metal cores can be dated with the $^{182}$Hf-$^{182}$W and the $^{53}$Mn-$^{53}$Cr chronometers on iron meteorites. A compilation of such ages for magmatic iron meteorites originating from the NC and CC reservoirs (Figure 4) yields a range in ages from ~0.4 Ma to 2.5 Ma after CAI formation for most irons (e.g., Kruijer et al., 2014; Spitzer et al., 2021). This makes some iron meteorites the second oldest material from the Solar System after CAIs. Iron meteorites from the NC reservoir yield core formation ages from ~0.4 Ma to 1.5 Ma, those from the CC reservoir are close to ~2.5 Ma after CAIs (e.g., Scott 2020; Anand et al., 2021b, Kruijer and Kleine, 2019; Kleine et al., 2005; Spitzer et al., 2021).

The oldest core formation ages predate the major formation interval of chondrules by 1.5 Ma to 2.0 Ma (Figure 3). These ages also demonstrate that differentiated planetesimals formed and differentiated within the first ca. 0.5 Ma of the Solar System. Their age range is consistent with $^{26}$Al as the main source of heating to temperatures for Fe-metal to melt. The complementary material to a metallic core is a silicate mantle. However, to date no achondritic material has been found among the known meteorites that is of similar antiquity as the oldest iron meteorites. Achondritic fragments have been found in chondrites, but they are of similar age as the host samples (e.g., Sokol et al. 2007). The oldest achondrite discovered to date is the meteorite Erg Chech 002, a coarse-grained andesite. It has been dated with multiple chronometers (e.g., Barrat et al., 2021; Anand et al., 2022) yielding a differentiation age of 1.83 ± 0.12 Ma after CAIs with the $^{26}$Al-$^{26}$Mg chronometer (Reger et al., 2023). This age overlaps with the initiation of chondrule formation.

The time of chondrite parent body accretion can be bracketed between the time of chondrule formation and the growth of new minerals in the chondrite parent body as a consequence of heating by decay of remaining $^{26}$Al. Dating chondrites from the same parent body, but with different metamorphic grades yielded a range of ages that correlate with peak metamorphic temperature of the samples (e.g., Trieloff et al., 2003, Anand et al., 2021a). The oldest ages obtained for samples with the lowest metamorphic grade imply an accretion earlier than ~3 Ma after CAIs (Anand et al., 2021a). The lower age limit is given by the ages of the youngest chondrules which formed at ~2.6 Ma after CAIs (Figure 3). Thus, the parent bodies of chondrites accreted immediately following chondrule formation (e.g., Yurimoto and Wasson, 2002; Ruzicka et al., 2024). Thus by ~3 Ma these parent bodies had formed. By that time most of the $^{26}$Al had decayed preventing large scale melting and core formation, and thus major planet-wide differentiation in chondrite parent bodies. As shown in Figure 3 the chondrules in ordinary chondrites are slightly older than those in carbonaceous chondrites. This difference may also apply to the time of accretion for ordinary vs. carbonaceous chondrite parent bodies. The metamorphic grade for ordinary chondrites reaches 7, whereas carbonaceous chondrites show only very low grades of 3 (except for the group CK). Thus, either the carbonaceous chondrite parent bodies were smaller than the parent bodies of ordinary chondrites, or accretion happened later when most $^{26}$Al had decayed and not enough radiogenic heat was produced to heat up the bodies to achieve higher metamorphic grades.

Ages from ~3.0-4.0 Ma after CAI formation are also recorded in other achondrites that derive from differentiated asteroids. A peculiar group of meteorites, i.e., ureilites, originate from a partially differentiated parent body that was likely still partially molten at the time of its early disruption (Rosén et al., 2019). Ureilites are rich in carbon (graphite, diamond; ~3 wt% on average; Warren and Huber, 2006) and show chemical and O-isotope compositions similar to carbonaceous chondrites (e.g., Clayton and Meyeda 1996; Kruttasch et al., 2025). However, their $^{50}$Ti and $^{54}$Cr compositions place them at the lower end of the NC reservoir (Warren 2011b; Kruttasch et al., 2025). Individual ureilite samples have homogeneous isotope compositions, but there are some variations in Cr, Ti and O isotope compositions between different meteorite samples (e.g., Kruttasch et al. 2025). The isotope heterogeneity and the arrested core-mantle differentiation places ureilites in a planetary evolution path between fully differentiated planetesimals that developed a metallic core and silicate mantle, and chondrite parent bodies. Based on $^{53}$Mn-$^{53}$Cr chronometry, differentiation of the ureilite parent body occurred at 2.9 $\pm$ 0.5 Ma and the body accreted no later than ∼1.5 Ma after CAI formation (Kruttasch et al., 2024). A similar age of 2.3 $\pm$ 0.9 Ma was obtained with the $^{182}$Hf-$^{182}$W chronometer for the oldest zircon in a basaltic eucrite that derive from the asteroid 4Vesta (Roszjar et al., 2016). These ages indicate that the ureilite and eucrite parent bodies differentiated at a time chondrite parent bodies accreted.

The combined chronology on different meteorites (e.g., Anand and Mezger, 2023) representing different stages of planetary accretion and differentiation indicates a protracted interval of planetesimal formation immediately following CAI formation that lasted up to ca. 4 Ma. Early formed planetesimals in the NC reservoir accreted and differentiated ~0.5 Ma before those in the CC reservoir. A similar time difference is indicated by chondrules from the two distinct reservoirs. The degree of differentiation of these planetesimals is a function of the time of accretion, as the radionuclide $^{26}$Al was the main heat sources early on, and the size of the bodies. Thus, chemically primitive materials from the Solar System do not indicate early formation, as late formation limited heating by short-lived isotopes and inhibited planet-wide differentiation into a metallic core and silicate mantle (e.g., Kunihiro et al., 2004). At ca. 4 Ma most of Mars had accreted (e.g., Dauphas and Pourmand, 2011).

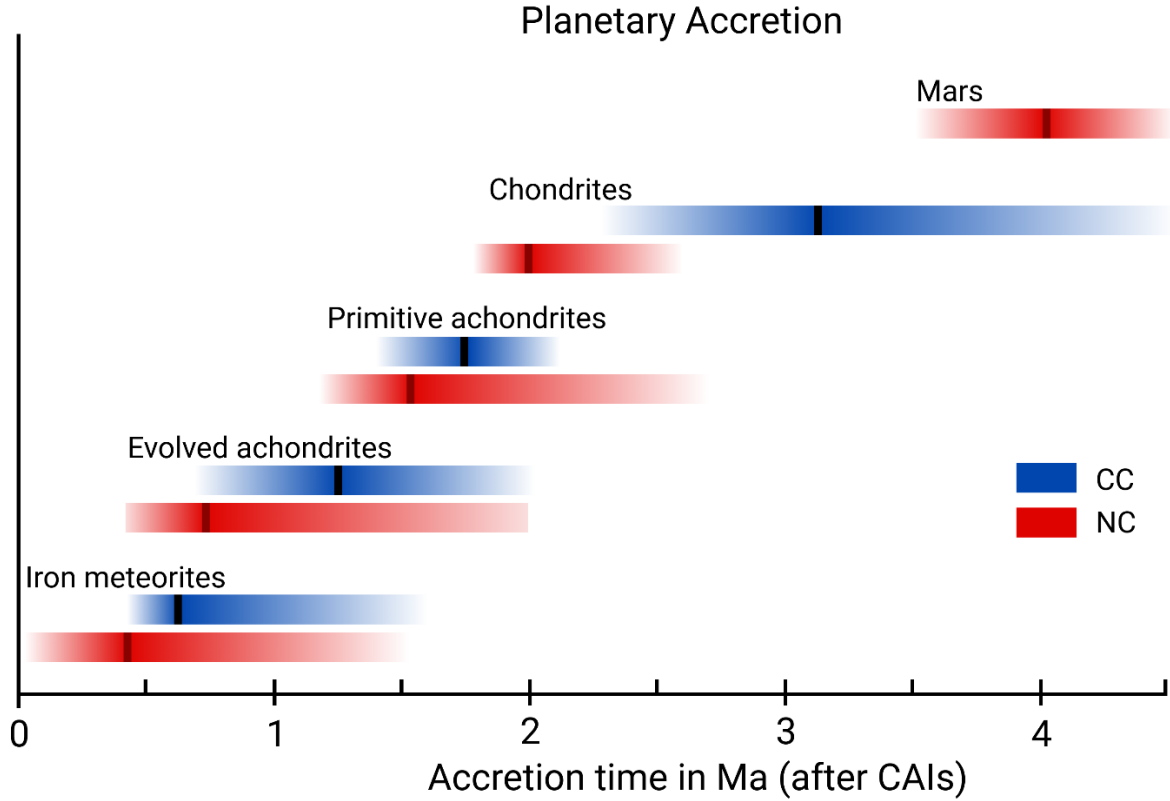

Fig 4: Time of accretion of planetesimals as recorded in meteorites. The times for accretion are deduced from major chemical fractionation events in the parent body, like formation of a metallic core, magmatism or metamorphic growth of new phases. Black vertical bars indicate age peaks.

**5 Earth-Moon system and Planetary Differentiation**

The highly variable chemical and isotope composition of different meteorites reflects the spatial and possibly temporal evolution of the solar nebula from the formation of the first dust grains to the accretion of planetesimals which became the building blocks of the known planets (e.g., Morbidelli et al., 2012). However, no extant meteorite samples, nor any combination of known meteorites representing a diversity of planetesimals, have a composition that is consistent with the bulk chemical and isotope composition of the Earth. The isotope compositions of refractory elements of enstatite chondrites and the Earth are very similar supporting an origin from the same reservoir. However, their bulk chemistry is quite different (e.g., Palme 2000) and so is the isotope composition of the volatile element Zn (Savage et al., 2022; Steller et al., 2022; Martins et al., 2023). The composition of s-process isotopes makes the Earth an endmember among the known samples from the Solar System (e.g., Mezger et al., 2020). Earth may contain material from known types of meteorites, but a significant component of unknown composition needs to be accounted for.

A partial reconstruction of the components that contributed to the Earth is possible from the composition of Bulk Silicate Earth (BSE). The element abundances in BSE also provide information on the chemical differentiation associated with the accretion of the Earth. In any accretion model the Moon has to be included as it has strong chemical and isotopic similarities with Earth.

The origins of Earth and Moon have to be considered together, due to the specific properties of these two planetary bodies that indicate a co-genetic origin. The Moon is strongly volatile element depleted, but the refractory elements have similar relative abundances compared to Earth. The nucleosynthetic isotope compositions of refractory elements and O (e.g., Akram and Schönbächler, 2016; Fischer et al., 2024) are analytically indistinguishable in both bodies. The Moon has a metallic core that is too small for a bulk chondritic composition of a planetary body. Thus, the Moon cannot have formed by accretion of primitive chondritic material like all other known planetary bodies in the Solar System. Relative to the size of the host planet, the Moon is the largest companion of any planet in the Solar System. The angular momentum of the Earth–Moon system is higher than for any other known terrestrial planet in the Solar System. These observations led to the currently favored model that the Moon formed by a later collision of a Mars-sized body with proto-Earth (e.g., Benz et al., 1986). Thus, present day Earth (and the Moon) is the product of the collision of two planetary bodies with possibly different compositions and evolution paths.

All the materials that ultimately contributed to the Earth originated from the solar nebula that was also the source of all the other planets and planetesimals in the Solar System. On the path from the first formation of solids to the final accretion of the Earth the building blocks can have undergone fractionation and mixing processes resulting in planetary bodies with different element abundances within the first ca. 4 Ma of the existence of the Solar System (Section 1). However, the composition of the Earth shows marked differences to primordial material as observed in the composition of the Sun (Asplund et al., 2009) or the most primitive meteorites represented by CI chondrites. In particular, the volatile elements show a strong depletion in Earth that is more pronounced than in chondrites (Section 2).

Different accretion models for Earth have been developed using known chemical and isotope data and inferred compositions of materials that could have contributed to the Earth and are commensurate with its composition (e.g., Wänke and Dreibus, 1988; O'Neill, 1991; Albarède, 2009; Schönbächler et al., 2010; Warren, 2011; Ballhaus et al., 2013; Rubie et al., 2015, 2016; Fitoussi et al. 2016; Liebske and Kahn, 2019). All these models take into account that the chemistry of the Earth cannot be the product of accretion of material that is identical to a single known meteorite type. It is possible to model some aspects of Earth`s chemistry by combining different known meteorite types (e.g., Rubie et al., 2015, 2016; Dauphas, 2017). However, the difficulty remains to account for the isotopic endmember composition of Earth indicated by the deficit in s-process isotopes (e.g., Mezger et al., 2020). One approach to explain the current bulk composition of the BSE is to combine multiple components with their characteristic compositions as building blocks with a specific sequence of accretion and differentiation. The model needs to account for the element abundances and isotope signatures of elements with distinct geochemical affinities in the different geochemical reservoirs of the Earth.

For the accretion model of Earth, it is assumed that the element abundances in BSE are the result of accretion of material with different extents of volatile element depletion and chemical differentiation processes at distinct stages during the assembly of the Earth. The element abundances in BSE normalized to the bulk Solar System composition (i.e., CI chondrite composition) and displayed versus their respective condensation temperature (Lodders, 2003; Wood et al., 2019) show different element groups that have relative chondritic abundances within the group (Figure 5). The three distinct groups are: a) refractory lithophile elements, b) moderately volatile lithophile elements, and c) highly siderophile elements. The stepped element pattern and the characteristic grouping of the elements imply that no single meteorite group can have been the sole source of the Earth. The pattern suggests three chemically, and possibly isotopically, different building blocks contributed to the final composition of the Earth (Wänke and Dreibus, 1988; Mezger et al., 2021). The strong depletion of the highly siderophile elements as well as the chalcophile elements, particularly S, Se and Te, in BSE (e.g., Wang and Becker, 2013) indicates removal of a metal melt and a sulfide-rich melt at some stage in the evolution of the Earth.

The composition of BSE can be modelled by combing three chemically distinct components. The major component of the Earth was highly depleted in volatile elements and had a low O-fugacity (Figure 5) (Wänke and Dreibus, 1988). This component (A; proto-Earth) is not represented by any known meteorite group, but its isotope composition has strong similarities to the NC reservoir. The second component (B; Theia) was less depleted in the highly volatile elements and more oxidized than the major component. Its isotope signatures indicate a strong affinity to the carbonaceous chondrite reservoir. A minor component (C; Late Veneer) with element abundances similar to CM chondrites (Figure 2) is recorded by the low abundances of platinum group elements in BSE. These three components needed to model the composition of the BSE have some similarities with different extant meteorite classes, but no single class fits with all chemical and isotopic parameters (Mezger et al., 2021).

Mixing these three components in a specific sequence and considering element fractionation by core formation, reproduces the element pattern of BSE (Figure 5). The element pattern of BSE is consistent with bulk Earth being a mixture of the two major components A and B in a ratio of 9:1. Considering only BSE (=bulk Earth without the metal core), the element abundances can be modelled by mixing the three components in the proportions 85:15:0.4. In order to reproduce the element pattern of BSE, these three components had to be mixed in a specific sequence. Component A evolved as a highly volatile element depleted and strongly reduced proto-Earth. This rocky planet was likely also dry (e.g., Gillmann et al., 2020; Salvador et al., 2023). $^{53}$Cr model ages constrain the time of element depletion in component A to the first 3 Ma of the Solar System (Kruttasch and Mezger., 2025). The accretion of reduced material resulted in efficient core formation depleting the silicate mantle in highly siderophile elements including most Fe, as it occurred as metallic Fe in this body. Subsequent addition of component B brought in material only depleted in the highly volatile elements and more oxidized than component A. This accretion step was followed by a second core formation event that involved segregating of a sulfide-melt from the silicate Earth (Hadean matte: O`Neill 1991; Rubie et al., 2016), which removed the siderophile elements almost entirely from the mantle together with S, Se and Te, the most depleted elements in BSE. This component B is attributed to Theia, the impactor that struck proto-Earth and caused the formation of the Moon. Following the second core formation, near-chondritic material was added as a late veneer to a silicate mantle that was almost completely devoid of the platinum group elements as well as S, Se and Te. Such a late veneer is consistent with the chondritic abundance of highly siderophile elements in BSE.

Using the chemical constraints derived from the element abundances in BSE and the model for Earth accretion with the U-Pb and Rb-Sr systematics of the Earth-Moon system yields an age of ~4.50 Ga for the collision of Theia with proto-Earth and the formation of the Moon (Maltese and Mezger, 2020; Mezger et al., 2021). This age postdates by about 70 Ma the frequent collisions recorded in planetesimals (Figure 6). In addition, the isotope composition of BSE shows a record of mixing of material originating from very different regions of the Solar System. Proto-Earth accreted from strongly volatile-depleted material of the inner Solar System supported by the isotope composition of refractory elements that are similar to enstatite chondrites and only marginally different from ordinary chondrites. Component B (=Theia) was less volatile element depleted and originated from a location at greater distance from the Sun. Since the moderately volatile elements in BSE originate dominantly from this component, their isotope composition should reflect this source.

The isotope composition of the volatile element Zn provides key information on the genealogy of the second component and its possible source region in the Solar System. The Zn isotope compositions of chondrites show a pronounced dichotomy; all samples analyzed so far belong either to the NC or the CC reservoir. However, Earth has an intermediate value which implies that the Zn in Earth derives from both the NC and CC reservoirs and ca. 30% of the Zn in BSE originates from the CC reservoir (Savage et al., 2022; Steller et al., 2022; Martins et al., 2023; Nimmo et al., 2024). This isotope signature places the origin of Theia at a much greater distance from the Sun than proto-Earth. Since most of the volatile element budget in Earth is derived from the impactor, Theia consisted of NC and CC material (e.g., Branco et al., 2025). This origin of Theia is also consistent with its higher volatile element content and stronger oxidized bulk composition compared to proto-Earth. However, no known meteorite sample fits exactly the composition of this hypothetical planetary body.

The combination of the chemical composition of BSE with the age constraints for the time of collision of Theia with proto-Earth supports an origin of Theia beyond Mars near the transition region of the NC and CC reservoirs. Since Mars has a Zn isotope composition characteristic of the NC reservoir and Earth is a mixture of material from NC and CC reservoirs, addition of small planetesimals originating from the outer regions to the rocky planets cannot account for the isotope signature, as Mars should have received material from the CC reservoir as well (Kleine et al., 2023). Theia exclusively added volatile rich material to the Earth. This makes the Giant Impact a singular and decisive event in the transformation of a rocky planet in the inner Solar System to a habitable planet with enough volatiles to sustain life. In addition, the Moon, which formed as a result of this collision, stabilized the rotation axis of the Earth through geological time permitting later biological evolution to progress on a planet with stable climate conditions. The presence of a metallic core that remained at least partially liquid created a magnetic field that protected life on the surface of the Earth from solar wind and cosmic radiation allowing life to emerge from the oceans onto the continents. This emphasizes the significance of the Moon-forming event for the evolution of advanced life on Earth.

The formation of the Earth-Moon system ca. 70 Ma after the beginning of the Solar System combined with the origins of Theia and proto-Earth at different locations in the Solar System implies a significant orbital instability in the planetary constellation at that time. Either Theia was in an unstable orbit, or it was moved into a position that caused it to migrate towards Earth, possibly by changes in the orbit of Jupiter (Walsh et al., 2011).

After the Giant Impact and after completion of core-formation, ca. 0.4% of chondritic material was added to Earth yielding the chondritic relative abundances of the highly siderophile elements in BSE (Figure 5). This component C delivered almost quantitatively the entire budget of highly siderophile elements currently residing in BSE. Since S, Te and Se show similar abundances in BSE (e.g., Wang and Becker, 2013), but are depleted relative to the highly siderophile elements, the element pattern of this late veneer was similar to CM chondrites or more volatile element depleted. Since this material was volatile element depleted, it was not the dominant source of volatile elements on Earth, supporting their origin from Theia that formed close to the CC reservoir.

The element and isotope budget of the Earth compared with that of different meteorite samples from a diversity of planetesimals provides key constrains on the material that contributed to Earth and the chemical differentiation associated with accretion. The final accretion of the Earth, its final major chemical differentiation into a chalcophile and siderophile element depleted mantle and a core, and the formation of the Moon are contemporaneous processes. This also implies that the habitability of the Earth is a direct result of the Giant Impact, a late chance event in the Solar System, which led to a mixture of strongly volatile depleted with undepleted material significantly after the major episode of planet formation.

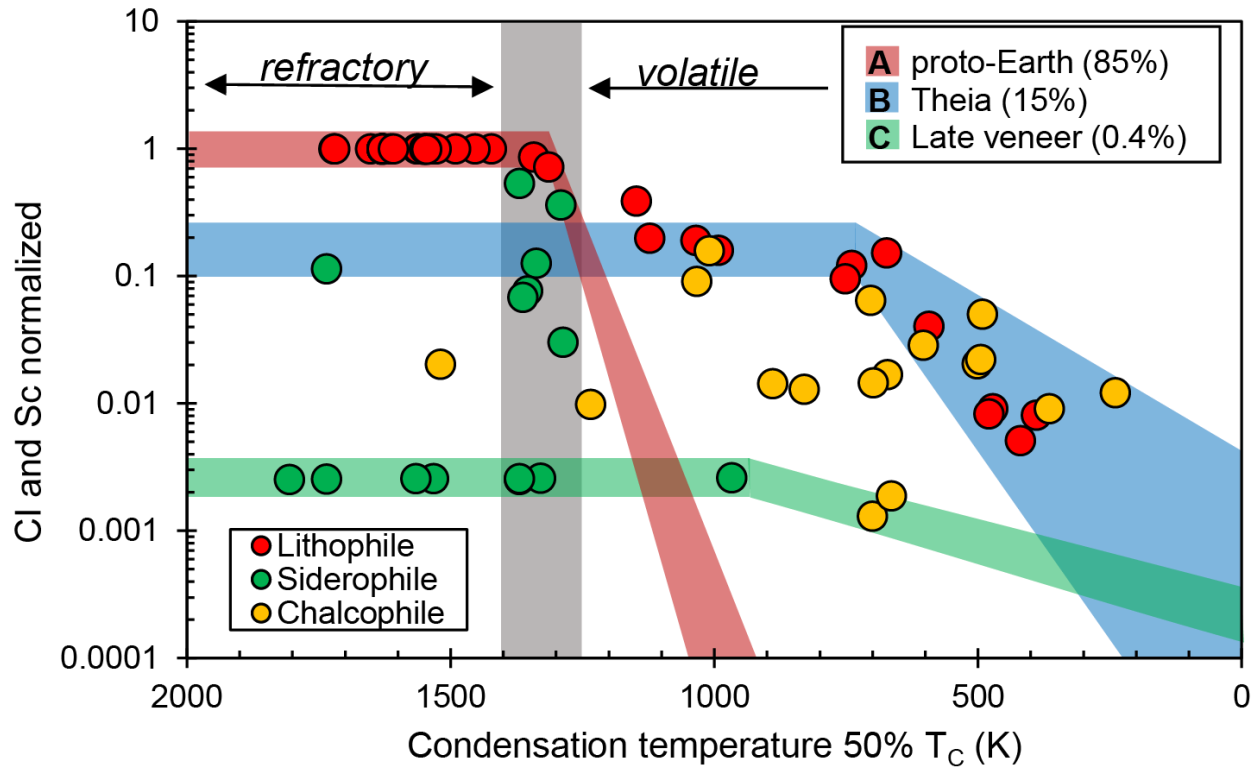

Figure 5: The element abundances in the silicate Earth can be modelled as the result of accretion of three distinct components with distinct element abundances and isotope composition: A: a strongly volatile element depleted and reduced proto-Earth, B: a moderately volatile depleted and more oxidized Moon-forming impactor (Theia) and C: a near chondritic and oxidized late veneer. The absolute abundance of the siderophile and chalcophile elements was modified by core formation following accretion (after Mezger et al., 2021).

## 6 Timeline of the early Solar System

The time and duration of many early Solar System processes can be derived from meteorites and their components using long and short-lived radio-isotope systems. Due to the rapid decay of short-lived isotopes, they can provide high temporal resolution for the first few Ma of the Solar System. Decay systems that yield precise and highly resolved age information include $^{26}$Al-$^{26}$Mg, $^{182}$Hf-$^{182}$W and $^{53}$Mn-$^{53}$Cr. Ages derived from these chronometers are based on the assumption that the radio-isotopes were homogeneously distributed in the solar nebula (Desch et al., 2023a, b). Due to the extinction of the parent isotopes, these systems can only yield absolute ages if the isotope information is anchored to an absolute age determined with the long-lived U-Pb system. Anchoring the short-lived $^{182}$Hf-$^{182}$W and $^{53}$Mn-$^{53}$Cr systems to the Pb-Pb age of the quenched angrite D'Orbigny (Tissot et al., 2017), yields an absolute age for CAIs of 4568.54 ± 0.19 Ma (Bouvier and Wadhwa, 2010; Kruttasch and Mezger, 2025; Kruttasch et al., 2024). CAIs formed in a short time interval of <0.1 Ma and being the oldest dated solids, they mark the time of "beginning of the Solar System". The ages in Figures 3 and 6 can be translated into absolute ages using this reference age.

The presence of short-lived isotopes, particularly the short-lived isotope $^{26}$Al, implies that these isotopes formed in a supernova just prior to their introduction into the solar nebula (Desch et al., 2022). Iizuka et al. (2025) estimated that the supernova that brought $^{26}$Al to the solar nebula exploded at 0.94 +0.25/-0.21 Ma before the condensation of the first solids. Due to the relatively high initial abundance of $^{26}$Al, the decay of this isotope can be used as a reliable chronometer for many early Solar System processes within the first ~3 Ma of the Solar System history. Moreover, $^{26}$Al was a major heat source in the early Solar System promoting early planetary differentiation by radiogenic heating of accreting planetesimals.

Chronometers based on radio-isotopes can yield precise and highly resolved ages for processes that involve significant fractionation of the parent from the daughter element. This makes different isotope systems suitable for dating particular processes, including the formation of the first solids from the solar nebula, formation of chondrules, differentiation of planetesimals into a metallic core and silicate mantle, metamorphism and melt production in planetesimals, heating and cooling of planetesimals and hydration of planetesimals. Dating these different events and processes reveals a complex evolution and a diversity of fractionation events and processes during the first ca. 10 Ma of the Solar System (e.g., Anand and Mezger, 2023).

The materials from the Solar System with the oldest known age are CAIs, fragments of solids dominated by highly refractory elements. These fragments are thought to have formed early in a hot environment, most likely close to the Sun. The next oldest materials are magmatic iron meteorites from parent bodies that formed dominantly in the NC reservoir, shortly followed by irons from the CC reservoir (e.g., Anand et al., 2021; Spitzer et al., 2021). This core formation was ongoing for several Ma with overlapping core formation between the two reservoirs. These iron meteorites represent the cores of differentiated planetesimals. Thus, the accretion of these planetesimals must predate differentiation. Compared to the formation of CAIs, planetesimal accretion (and differentiation) started immediately after the formation of an accretion disk following the condensation of the highly abundant elements, including Si, Fe and Mg, into solids. Planetesimal accretion was an ongoing process that occurred in different places of the solar nebular over at least 5 Ma (Figure 6). These accretion events do not correlate with heliocentric distance.

This first phase of planetesimal formation also created the major compositional gap in the Solar System at the location where Jupiter accreted (e.g., Kruijer et al., 2017, Brasser and Mojzsis, 2020) within the first 1 Ma. These early planets and planetesimals may have been the cause for the dissection of the solar nebula into regions with early formed planetesimals and regions made up of dust and gas, which resulted in a ringed structure of the solar nebula that was similar to nascent planetary systems imaged by ALMA (e.g., Flock et al., 2015). Moreover, this separation into distinct reservoirs with different chemical and isotopic bulk compositions (e.g., NC-CC; Figure 1) led to the formation of planetesimals and planets with their characteristic isotope compositions (e.g., Rüfenacht et al., 2023; Kruijer et al., 2017) and attests to limited material transport across the nebular disk once the first gaps had formed. The development of this major discontinuity that led to the separation of the NC from the CC reservoir (Figure 1) has been attributed to the formation of Jupiter (Kruijer et al., 2017), the development of a pressure maximum in the disc near the location of Jupiter (Brasser and Mojzsis, 2020), or the migration of the snowline (Lichtenberg et al., 2021). However, it is not resolved, if this dichotomy of an NC and a CC reservoir is a primary feature of the solar nebula or the result of a gap forming around the location of Jupiter. Accretion of material in this location from a gas-dust cloud with a continuous gradient in isotope compositions can also have created an apparent discontinuity in the isotope composition of the solids.

The first interval of planetary accretion and differentiation was followed by the formation of chondrules probably by melting of dust aggregates. Chondrule formation commenced at ca. 1.8 Ma and ended at <3 Ma. Chondrules from group CB and CH formed even later, but likely by a different process (e.g., Krot et al., 2005). The process involved during the major chondrule forming interval involved flash heating followed by rapid cooling of the once molten chondrules. This heating event occurred multiple times as shown by the age differences in the main chondrule forming episode in the NC and CC reservoirs (Figure 3) and the observation of multiple melting events in single chondrules (e.g., Pape et al., 2021). The chondrules were incorporated into planetesimals immediately after their genesis (e.g., Anand et al., 2021a, Trieloff et al., 2003). These second generation planetesimals did not differentiate into core and mantle due to their late formation, at the time when most $^{26}$Al had already decayed. This allowed the preservation of the chemically primitive composition of the chondrite parent bodies.

Some meteorite groups show evidence of hydration, particularly the CM and CI chondrites. They also have the least depletion in volatile elements and may be the accessible material that formed most distantly from the Sun, where ice was accreted together with rocky material. Heating of these primitive bodies led to the formation of hydrous minerals. This metamorphism occurred later than ca. 4.5 Ma after CAIs (e.g., Fujiya et al., 2013). At this time Mars had already mostly accreted (e.g., Dauphas and Pourmand, 2011).

The combined chronological information on early Solar System processes demonstrates that accretion and differentiation of planetary bodies did not follow a unique timeline but was a protracted process that occurred diachronous across the solar disk. The formation of the first planetesimals was early enough to cause planet-wide melting and differentiation into a metallic core and a silicate mantle. The place within the disk where accretion occurred was random, supporting the process of streaming instability as the cause for accumulation of solids into planetesimals (e.g., Helled et al., 2025). The existence of the later formed chondrite parent bodies with their chemically primitive composition indicates that the regions of dust and gas between the orbits of accreted planetesimals remained primitive and there was only minor exchange of material across the ring structure of the evolving solar disk.

The accretion of individual planetary bodies was a local phenomenon with stochastic spatial distribution. The accretion of planetesimals was a protracted process that lasted up to ca. 4 Ma and coincided with the formation of Mars and possibility proto-Earth.

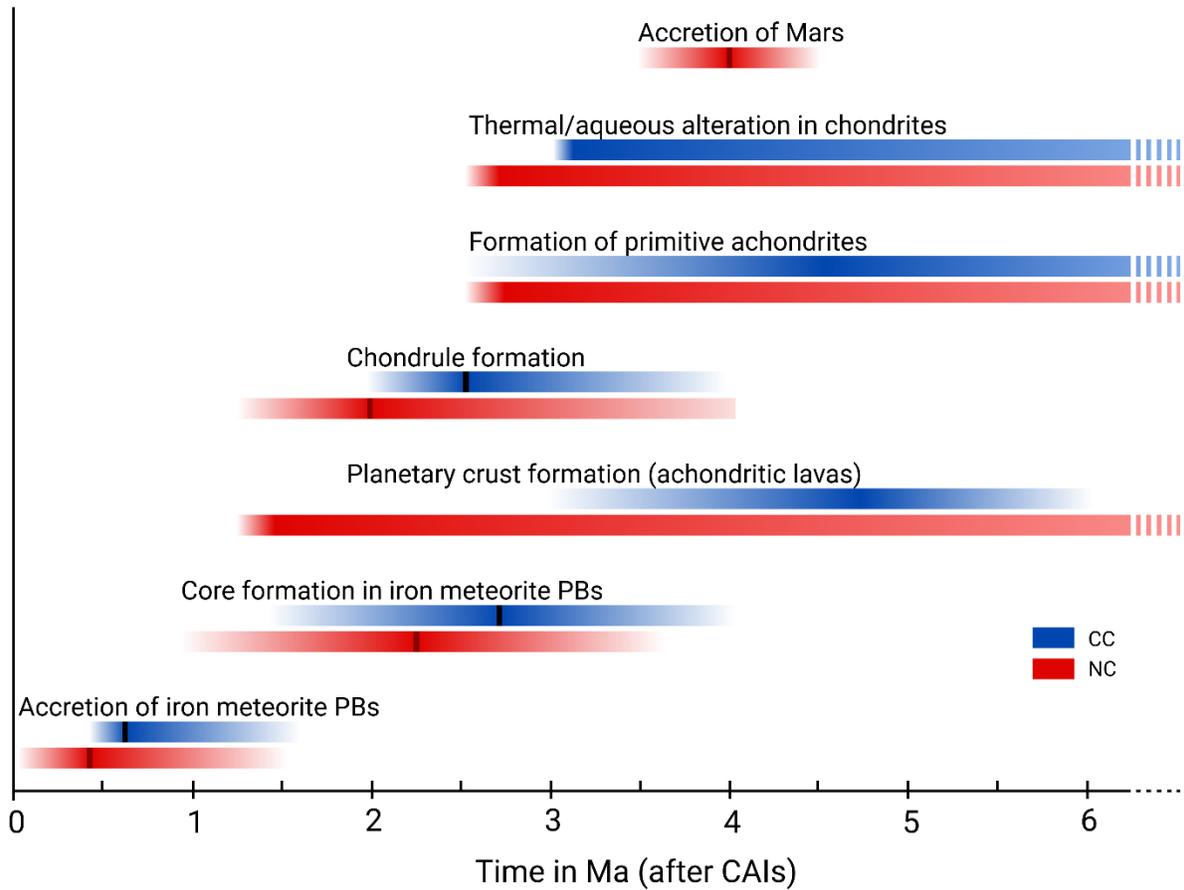

Figure 6. Timeline of key processes and events in the early Solar System from the formation of the first solids (CAI) to the accretion of planets. Black vertical bars indicate age peaks.

# 7 Conclusions

The chemical, mineralogical and isotopically diverse rocky materials from the Solar System available for direct study in the laboratory preserve a record for the formation of solids and their accretion into planetesimals as a function of time. The diversity of planetary bodies in terms of size, composition and heliocentric distance in the Solar System implies different paths in space and time for physical and chemical processes in a single Solar System. The timing of events and processes from the formation of the first solids to accretion of planetesimals and planets and their chemical differentiation reveals a complex temporal and spatial evolution of the early Solar System. Intervals of steady evolution (i.e., cooling, condensation, contraction of gas-dust cloud) were interrupted by singular events (i.e., bow shocks, chondrule formation, accretion of first planetesimals, planet migration, Giant Impact) that changed the direction of evolution. These observations deduced from the analysis of different materials from our Solar System can be a test for any universal model describing the formation of planetary systems in general. Any universally applicable model needs to take into consideration singular events that are turning points that can change a steady evolution and give it a new direction. Meteorites and their components provide key constraints on important singular events and turning points in the early evolution of the Solar System that led to a rocky planet like the Earth in the habitable zone capable of supporting life. Several singular events were also responsible for enabling the evolution of complex life on Earth.

**Short-lived radio-isotopes:** The majority of the chemical elements contained in solar nebula derived from at least two distinct stellar sources (Figure 1). Minor contributions can be attributed to several more nucleosynthetic sources (e.g., Hoppe 2008). At least one of these major sources contained short-lived radio-isotopes indicating the synthesis of this material just prior to the collapse of the parental molecular cloud (e.g., Izuka et al., 2025). Particularly the $^{26}$Al was a dominant heat-source during the first few Ma of the Solar System. Early formed planetary bodies were thermally modified by radiogenic heat from the decay of $^{26}$Al, augmented by impact heating, leading to partial melting and segregation of a metallic core. Thus, all early formed bodies differentiated into two chemically very distinct reservoirs, i.e., a metallic core and a silicate mantle. A lower amount of $^{26}$Al or its absence during the initial stages of Solar System evolution would have limited core formation and thus the major chemical differentiation event in planetary bodies.

A higher amount of $^{26}$Al would have led to differentiation of late formed planetesimals like the parent bodies of carbonaceous chondrites. For the early stages of planetary evolution, the timing of the injection of $^{26}$Al and its amount into the solar nebula were critical, particularly for early chemical differentiation of rocky planetesimals and planets. These values are not predictable and need to be considered as variables in models for the chemical and physical evolution of planetary systems.

The storage of reduced Fe in the Earth`s core was important for the evolution of multicellular life, as early life needed to create an $O_2$-bearing atmosphere, which was only possible after oxidation of Fe on land surfaces and in the ocean. Due to the removal of reduced Fe to the core, an $O_2$-bearing atmosphere could be achieved by ca. 2.4 Ga on Earth (first Great Oxygenation Event; e.g., Farquhar et al., 2000), enabling the evolution of multicellular life. The presence of $^{26}$Al as a source of thermal energy in the early Solar System played a key role in the formation of rocky differentiated planets and the evolution of Earth towards a habitable planet capable of supporting complex multicellular life

**Isotope dichotomy in the early Solar System:** The dynamic evolution of the solar nebula led to the formation of the first planetary bodies within ca. 0.5 Ma of the Solar System (Figure 6). At about this time a major discontinuity formed that was either caused by the accretion of Jupiter (Kruijer et al., 2017), the formation a pressure-maximum in the region where Jupiter is located (Brasser and Mojzsis, 2020), or the migration of the snowline (Lichtenberg et al., 2021). Mass transport across this discontinuity was largely inhibited, but most likely not completely. This led to a separation of the solar nebula into two distinct isotope reservoirs, the NC- and the CC-reservoir. The site of this discontinuity was probably due to a random occurrence of turbulence that led to rapid growth of Jupiter, a planet that was big enough early and able to attract H and He, resulting in the largest planet of the Solar System. This planet may be the reason for the small size of Mars. Over geological time the presence of a giant gas planet with its strong gravitational field limited the possibility for the delivery of volatile-rich material from the outer region of the Solar System to the volatile element-poor rocky planets of the inner solar. Thus, the architecture of the Solar System was strongly influenced by the early formation of a discontinuity in a chance region of the cooling gas-dust cloud.

**Formation of the first planetesimals:** The first planetesimals, and possibly even Jupiter, formed by concentration of solids due to turbulence in the gas-dust cloud of the early Solar System. The accretion of the first generation of planetary bodies occurred in random regions where turbulence increased the dust/gas ratio (Helled et al., 2025; Johansen et al., 2007) producing a ring structure, separating regions composed of dust and gas from regions where this material was concentrated in planetesimals or planets. This ring structure limited lateral element transport and mixing of different materials within the disk. As a consequence, each planetary body acquired a distinct isotope composition. These early formed bodies were the likely cause for the formation of chondrules in the remaining dense gas-dust regions. The formation of chondrules was a likely trigger for the formation of a second generation of planetary bodies by pebble accretion that was fed by the local ring structure which placed also a limit on the size of the accreting bodies.

**Giant Impact:** The Earth accreted originally material from inside the snow-line resulting in a rocky planet that was strongly reduced and lacked volatile elements. The transition to a habitable planet is the result of a Mars-sized body striking proto-Earth and merging with it. The Giant Impact was the final major accretion step that shaped the Earth. This event is associated with a planet-wide chemical differentiation into a chalcophile and siderophile element depleted mantle and a metallic core, and the formation of the Moon.

Since rocky material and volatile rich material seem to exclude each other, (planets beyond the ice-line develop into giant gas planets) a mechanism for the delivery of volatile material is needed. In case of the Earth this was an accidental impact by the volatile-rich planet Theia. The most likely source for the highly volatile elements, including water, on Earth is the Moon-forming impactor. The habitability of Earth and its ability to develop plate tectonic processes is the result of this chance collision, because it added volatiles to a strongly volatile element depleted proto-Earth ca. 70 Ma after CAI, i.e., significantly after the major episode of planet formation. The higher concentrations of volatile elements in Theia indicate that this planetary body originated from beyond the orbit where Earth formed.

The late collision of planetary bodies with spatially separated origins also indicates significant instability in the orbits of the planets for a protracted time. The habitability of the Earth is thus a direct result of the Giant Impact; a chance event in the early Solar System, which was a consequence of Theia forming in an unstable orbit or its diversion from a stable orbit caused by another migrating planet.

**Formation of the Moon:** Planets can acquire moons through capture of stray asteroids or they can form from the debris produced by impacts onto their host planet. The latter is the most like case for Earth`s moon. Compared to other moons in the Solar System, the Earth has the largest moon relative to its size. The size and existence of the Moon play a major role in stabilizing Earth`s rotation and thus Earth`s long-term climate stability, which in turn promoted biological evolution and allowed the Earth to bring forth advanced forms of life.

The timeline of early Solar System processes (Figure 6) reveals a complex spatial and temporal evolution of the Solar System from the formation of the first solids to the accretion and differentiation of asteroids and planets in the present-day Solar System. Placing different events and processes of the early Solar System on a highly resolved timeline using meteorites and their components suggests that the path towards a habitable planet like the Earth, was set already early during the nebular stage. The subsequent thermal regime was controlled by short-lived isotopes and the formation of the first planetesimals was initiated in random places in the disk.

The specific observations deduced from meteorites and their components from our own Solar System suggest that the derivation of a uniform or "standard model" for the planetary system starting from a molecular cloud must take into account singular and random events and punctuated processes. The evolution of the Solar System followed intervals with gradual changes of physical and chemical parameters (e.g., mixing material from different nucleosynthetic sources in the nebula, cooling, accretion of dust) that led to "tipping points" that initiated a different phase of gradual change (e.g., condensation of major elements Fe, Mg and Si dramatically changed the gas-dust ratio and the associated physical parameters of the nebula). Other turning points in the evolutionary path were the result of singular events with a random occurrence (e.g., local streaming instability leading to planetesimal accretion, Giant Impact). The specific observations from our own Solar System suggest that any unified or general model for the formation of a planetary system with a habitable planet will be an "ideal model". Deviations from the model will have to be evaluated at every step and incorporated to derive a robust and realistic model for any given planetary system.

The laboratory study of a diversity of Solar System material has led to detailed insights into origin and evolution of one planetary system. However, some major questions remain for a complete reconstruction of the genesis of the Solar System that may have also implications for an understanding of the evolutionary path of multi- planetary systems in general.

Although materials from different Solar System objects show a diversity of isotope compositions that indicate a chemical gradient in the solar nebula, the spatial distribution of this gradient remains elusive. Combined with the observation that the Earth is a chemical endmember it would be of great importance to obtain material from the inner solar system, inside the orbit of Earth. Rocky samples from Venus and Mercury could complete the picture and help to reconstruct the chemical diversity of the solar nebula prior to accretion of planetesimals and planets.

Physical samples from distant bodies, including comets, may be necessary to explain the composition of gases in the atmospheres of the rocky planets. They could also provide primitive organic molecules that may have served as building blocks of life. This will help to understand if life can form on a rocky planet like the Earth and whether extraterrestrial material is needed or not.

The Giant Impact was probably a most decisive event in the formation of present-day Earth and the Moon. To constrain the timing of this event and to identify the composition and origin of Theia pristine samples from the lunar mantle could provide new and different information. Particularly subsurface samples would be highly desirable as they could be unaffected by modifications from later impacts on the Moon and long exposure to the solar wind and galactic irradiation. Most lunar samples were collected during the Apollo missions, and this collection seems to be biased as the samples returned by the Chang'e mission show (e.g., Zhang et al., 2025). The diversity of lunar rocks still needs to be explored.

Most meteorites are derived from planetesimals, but few can be related to a currently existing and observable body. Samples collected from the asteroids Ryugu and Benno enhanced our understanding of primitive, i.e., CI-like planetesimals. In-situ, or ideally sample-return missions from different asteroids could complete the catalog of diverse ancient planetesimals and allow a correlation of different meteorite classes to planetesimals. Such studies could also help to quantify the abundance and distribution of chemically different and distinct planetary bodies in the Solar System and relate them the abundance of the different meteorite groups in terrestrial collections, which may not be representative for the diversity of planetesimals.

A more complete reconstruction of the processes, events and chemical diversity in our Solar System will help to refine models describing the origin and evolution of planetary systems in general and reveal the generality of processes and significance of unique events.

*Acknowledgements:* The work presented in this Chapter has been mostly carried out within the framework of the NCCR PlanetS. Interactions and collaborations with researchers within and outside NCCR PlanetS were instrumental in the development, execution and completion of diverse research projects on meteorites. NCCR PlanetS made the acquisition of a Multicollector-ICP-Mass-Spectrometer possible, which enabled the acquisition of high precision isotope ratio measurements that became key elements of several research projects.